\documentclass[aps,twocolumn,floatfix,showpacs,eqsecnum,prb]{revtex4}
\usepackage{graphicx}
\usepackage{amsmath}
\usepackage{amssymb}
\usepackage{bm}

\begin{document}

\title{Random-matrix theory of Andreev reflection from a topological superconductor}
\author{C. W. J. Beenakker}
\affiliation{Instituut-Lorentz, Universiteit Leiden, P.O. Box 9506, 2300 RA Leiden, The Netherlands}
\author{J. P. Dahlhaus}
\affiliation{Instituut-Lorentz, Universiteit Leiden, P.O. Box 9506, 2300 RA Leiden, The Netherlands}
\author{M. Wimmer}
\affiliation{Instituut-Lorentz, Universiteit Leiden, P.O. Box 9506, 2300 RA Leiden, The Netherlands}
\author{A. R. Akhmerov}
\affiliation{Instituut-Lorentz, Universiteit Leiden, P.O. Box 9506, 2300 RA Leiden, The Netherlands}
\date{December 2010}
\begin{abstract}
We calculate the probability distribution of the Andreev reflection eigenvalues $R_{n}$ at the Fermi level in the circular ensemble of random-matrix theory. Without spin-rotation symmetry, the statistics of the electrical conductance $G$ depends on the topological quantum number $Q$ of the superconductor. We show that this dependence is nonperturbative in the number $N$ of scattering channels, by proving that the $p$-th cumulant of $G$ is independent of $Q$ for $p<N/d$ (with $d=2$ or $d=1$ in the presence or in the absence of time-reversal symmetry). A large-$N$ effect such as weak localization cannot, therefore, probe the topological quantum number. For small $N$ we calculate the full distribution $P(G)$ of the conductance and find qualitative differences in the topologically trivial and nontrivial phases.
\end{abstract}
\maketitle

\section{Introduction}
\label{intro}

The random-matrix theory (RMT) of quantum transport studies the statistical distribution of phase coherent transport properties in an ensemble of random scattering matrices. The theory finds a major application in the context of chaotic scattering, because then the ensemble is fully specified by fundamental symmetries --- without requiring microscopic input.\cite{Blu90} Since scattering phase shifts for chaotic scattering are uniformly distributed on the unit circle, such ensembles are called ``circular'', following Dyson who first introduced these ensembles in a mathematical context.\cite{Dys62} The circular ensembles have been successful in describing experiments on low-temperature electrical and thermal conduction in quantum dots, which are confined geometries connected by point contacts to metallic or superconducting electrodes. For a recent overview of the field we refer to several chapters of a forthcoming Handbook.\cite{RMTbook}

While metallic quantum dots are characterized by the three circular ensembles introduced originally by Dyson,\cite{Dys62} superconducting quantum dots are described by four different ensembles discovered by Altland and Zirnbauer.\cite{Alt97} The classification of the superconducting ensembles is based on the presence or absence of time-reversal and spin-rotation symmetry, as summarized in Table \ref{tab:table2}. The symmetry classes are called D, DIII, C, and CI, in a notation which originates from differential geometry.\cite{Alt97} The corresponding circular ensembles, in the nomenclature of Ref.\ \onlinecite{Dah10}, are the circular real (CRE) and circular quaternion (CQE) ensembles in class D and C, respectively, and their time-reversal invariant restrictions (T-CRE and T-CQE) in class DIII and CI. 

\begin{table*}
\centering
\begin{tabular}{ | l || c | c | c | c |}
\hline
Symmetry class\ &  D &  DIII & C & CI \\ \hline
Ensemble name & CRE & T-CRE & CQE & T-CQE \\ \hline
Particle-hole symmetry & \multicolumn{2}{|c|}{$r_{ee}=r_{hh}^{\ast}$, $r_{eh}=r_{he}^{\ast}$} &  \multicolumn{2}{|c|}{$r_{ee}=r_{hh}^{\ast}$, $r_{eh}=-r_{he}^{\ast}$} \\ \hline
Time-reversal symmetry & $\times$ & $r=\Sigma_{y}r^{\rm T}\Sigma_{y}$ &  $\times$ & $r=r^{\rm T}$ \\ \hline
Spin-rotation symmetry &  $\times$ &  $\times$ & \checkmark & \checkmark \\ \hline \hline
topological quantum number $Q$&${\rm Det}\,r$&${\rm Pf}\,i\Sigma_{y}r$&$\times$&$\times$ \\ \hline
degeneracy of $R_{n}\neq 0,1$  &  $2$ & $2$ & $2$ & $2$ \\ \hline
degeneracy of $R_{n}= 0,1$  &  $1$ & $2$ & $2$ & $2$ \\ \hline
\end{tabular}
\caption{Classification of the symmetries of the reflection matrix $r$ for a normal-metal--superconductor junction. See Sec.\ \ref{Andreevreflection} for explanations.}
\label{tab:table2}
\end{table*}

\begin{figure}[tb]
\centerline{\includegraphics[width=0.8\linewidth]{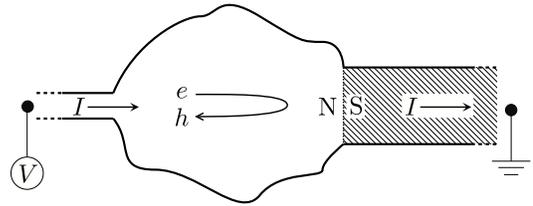}}
\caption{\label{fig_QDNS}
Quantum dot with an interface between a normal metal (N) and a superconductor (S, shaded region). Andreev reflection at the NS interface converts a normal current (carried by electron and hole excitations $e$ and $h$) into a supercurrent (carried by Cooper pairs). The conductance $G$ is the ratio of the current $I$ into the grounded superconductor and the voltage $V$ applied to the quantum dot via an $N$-mode point contact to a normal metal electrode (narrow opening at the left). The system is in a topologically nontrivial state if it supports a quasi-bound state at the Fermi level. This is possible if spin-rotation symmetry is broken by spin-orbit coupling. In the configuration shown in the figure (with a single NS interface), time-reversal symmetry should be broken to prevent the opening of an excitation gap in the quantum dot. In the presence of time-reversal symmetry a second NS interface, with a $\pi$ phase difference, can be used to close the gap.
}
\end{figure}

In a remarkable recent development, it was found that some of these symmetry classes divide into disjunct subclasses, characterized by a topological quantum number.\cite{Kit01,Qi08,Sch08,Fu09} For a quantum dot connected to a superconducting wire, such as shown in Fig.\ \ref{fig_QDNS}, the topological quantum number $Q=-1$ or $+1$ depending on whether or not the quantum dot has a bound state at zero excitation energy. (The state is only quasi-bound if the quantum dot is also connected by a point contact to a normal-metal electrode.) Because of particle-hole symmetry, such a fermionic excitation is equal to its own antiparticle, so it is a Majorana fermion. There is now an active search for the Majorana bound states predicted to appear at the ends of superconducting wires without spin-rotation symmetry.\cite{Lut10a,Ore10,Wim10,Pot10,Lut10b,Duc10}

The RMT of the four superconducting circular ensembles was developed for the quasiparticle transmission eigenvalues in Ref.\ \onlinecite{Dah10}, and applied to the thermal conductance. The probability distribution of this transport property does not depend on the topological quantum number, so it was not needed in that study to distinguish the topologically nontrivial ensemble (with a Majorana bound state) from the topologically trivial ensemble (without such a state).

The electrical conductance $G$, in contrast, can probe the presence or absence of the Majorana bound state through resonant Andreev reflection.\cite{Law09,Fle10} This applies also to a quasi-bound state,\cite{Ber09b,Wim11} so even if the $N$-mode point contact to the normal metal has a conductance which is not small compared to $e^{2}/h$ --- although the effect of $Q$ on $G$ diminishes with increasing $N$. As we will show in this paper, the $Q$-dependence of the conductance distribution $P(G)$ in the circular ensembles is nonperturbative in $N$: Cumulants of order $p$ are identical in the topologically trivial and nontrivial phases for $p<N$ ($N/2$) in the absence (presence) of time-reversal symmetry.

We derive this nonperturbative result by exactly computing (in Sec.\ \ref{RMT} and App.\ \ref{AppA}) the joint probability distribution $P(\{R_{n}\})$ of the Andreev reflection eigenvalues $R_{1},R_{2},\ldots R_{N}$ (eigenvalues of the product $r_{he}^{\dagger}r_{he}^{\vphantom{\dagger}}$ of the matrix $r_{he}$ of Andreev reflection amplitudes). This probability distribution was only known previously for one of the ensembles (CQE) without a topological phase,\cite{Kha10} while here we calculate it for all four superconducting ensembles, including the possibility of a topologically nontrivial phase. 

The distribution of the electrical conductance $G\propto\sum_{n}R_{n}$ follows upon integration over the $R_{n}$'s (Sec.\ \ref{conductance}). For small $N$ we obtain a closed-form expression for $P(G)$ in the two cases $Q=\pm 1$, and we demonstrate that, upon increasing $N$, first the average becomes the same, followed by variance, skewness, kurtosis, etc. A proof for arbitrary $N$ of the $Q$-independence of low-order cumulants is given in App.\ \ref{AppB}.

We conclude in Sec.\ \ref{conclude} with a comparison of the analytical predictions from RMT with a computer simulation of a system that is of current experimental interest (a disordered semiconductor wire on an $s$-wave superconducting substrate, with Rashba spin-orbit coupling and Zeeman spin splitting).\cite{Lut10a,Ore10}

\section{Andreev reflection eigenvalues}
\label{Andreevreflection}

Andreev reflection of electrons injected through a point contact towards a normal-metal--superconductor (NS) interface is described by the $N\times N$ subblock $r_{he}$ of the $2N\times 2N$ reflection matrix $r$,
\begin{equation}
r=\begin{pmatrix}
r_{ee}&r_{eh}\\
r_{he}&r_{hh}
\end{pmatrix}.\label{rblock}
\end{equation}
The labels $e$ and $h$ refer to electron and hole excitations in the normal-metal point contact, each of which can be in one of $N$ modes. We allow for spin-orbit coupling, so $N$ counts both orbital and spin degrees of freedom. The Andreev reflection eigenvalues $R_{n}$ ($n=1,2,\ldots N$) are the eigenvalues of the Hermitian matrix product $r_{he}^{\dagger}r_{he}^{\vphantom{\dagger}}$.

At excitation energies $\varepsilon$ below the superconducting gap $\Delta$ there is no transmission of excitations through the superconductor (assuming that it is large compared to the penetration depth $\xi_{0}=\hbar v_{F}/\Delta$). The reflection matrix is then a unitary matrix, $r^{\dagger}=r^{-1}$. An electrical current $I$ can still enter into the superconductor, driven by a voltage difference $V$ with the normal metal. The electrical conductance $G=I/V$ is fully determined by the Andreev reflection eigenvalues,\cite{Blo82}
\begin{equation}
G/G_{0}=N-{\rm Tr}\,r_{ee}^{\dagger}r_{ee}^{\vphantom{\dagger}}+{\rm Tr}\,r_{he}^{\dagger}r_{he}^{\vphantom{\dagger}}=2\sum_{n=1}^{N}R_{n}.\label{BTK}
\end{equation}
The conductance quantum is $G_{0}=e^{2}/h$ and the factor of two accounts for the fact that charge is added to the superconductor as Cooper pairs of charge $2e$. (Spin is counted in the sum over $n$.)

The relation \eqref{BTK} holds in the limit of zero voltage and zero temperature, when the reflection matrix is evaluated at the Fermi level ($\varepsilon\rightarrow 0$). The subblocks in Eq.\ \eqref{rblock} are then related by particle-hole symmetry,
\begin{equation}
r_{hh}^{\vphantom{\ast}}=r_{ee}^{\ast},\;\;r_{eh}^{\vphantom{\ast}}=r_{he}^{\ast}.\label{ehsymmetry}
\end{equation}
Time-reversal symmetry imposes a further constraint on the reflection matrix,
\begin{equation}
r=\Sigma_{y}r^{\rm T}\Sigma_{y},\label{rSigmayrT}
\end{equation}
with $\Sigma_{y}=\sigma_{y}\oplus\sigma_{y}\oplus\cdots\oplus\sigma_{y}$ and $\sigma_{y}$ a $2\times 2$ Pauli matrix. (The superscript T denotes the transpose.)

The Andreev reflection eigenvalues are all twofold degenerate in the presence of time-reversal symmetry. This is the usual Kramers degeneracy, which follows directly from the fact that $\Sigma_{y}r$ is an antisymmetric matrix [$\Sigma_{y}r=-(\Sigma_{y}r)^{\rm T}$] when Eq.\ \eqref{rSigmayrT} holds.\cite{Bar08} 

Remarkably enough, a twofold degeneracy persists even if time-reversal symmetry is broken. More precisely, as discovered by B\'{e}ri,\cite{Ber09a} if $R_{n}$ is not degenerate then it must equal $0$ or $1$. This follows from the antisymmetry of the matrix $r_{he}^{\rm T}r_{ee}^{\vphantom{t}}$, which is required by particle-hole symmetry and unitarity irrespective of whether time-reversal symmetry is present or not. The degeneracy of the Andreev reflection eigenvalues $R_{n}\neq 0,1$ is remarkable because the eigenvalues of the Hamiltonian are not degenerate in the absence of time-reversal symmetry. To distinguish it from the Kramers degeneracy, we propose the name ``B\'{e}ri degeneracy''.

The determinant of $r$ is real due to particle-hole symmetry, and therefore equal to either $+1$ or $-1$ due to unitarity. The topological quantum number $Q={\rm Det}\,r$ distinguishes the topologically trivial phase ($Q=1$) from the topologically nontrivial phase ($Q=-1$).\cite{Boc00,Mer02,Akh10} This quantum number is ineffective in the presence of time-reversal symmetry, when Kramers degeneracy enforces ${\rm Det}\,r =1$. In that case the Pfaffian (the square root of the determinant of an antisymmetric matrix) can be used instead of the determinant to identify the topologically nontrivial phase:\cite{Ful11} $Q={\rm Pf}\,i\Sigma_{y}r$ equals $+1$ or $-1$ depending on whether the superconductor is topologically trivial or not. 

A topologically nontrivial superconductor has a (possibly degenerate) bound state at $\varepsilon=0$, consisting of an equal-weight superposition of electrons and holes from the same spin band. It is the $\pi$ phase shift upon reflection from such a Majorana bound state which is responsible for the minus sign in the topological quantum number.\cite{Akh10}

These properties of the reflection matrix in the absence of spin-rotation symmetry are summarized in Table \ref{tab:table2}. For completeness, we also include in that table the case when there is no spin-orbit coupling. In that case it is sufficient to consider only the orbital degree of freedom, with a two-fold spin degeneracy of all $R_{n}$'s. The conditions for particle-hole symmetry and time-reversal symmetry are then both different from Eqs.\ \eqref{ehsymmetry} and \eqref{rSigmayrT}, given respectively by
\begin{align}
&r_{hh}^{\vphantom{\ast}}=r_{ee}^{\ast},\;\;r_{eh}^{\vphantom{\ast}}=-r_{he}^{\ast},\label{ehsymmetry2}\\
&r=r^{\rm T}.\label{TRSsymmetry}
\end{align}
As a consequence, the determinant of $r$ is now always $+1$, while the Pfaffian does not exist (for want of an antisymmetric matrix). Without spin-orbit coupling Andreev reflection exclusively couples electrons and holes from opposite spin bands, which prevents the formation of a Majorana bound state at the NS interface.

\section{Random-matrix theory}
\label{RMT}

In this section we calculate the distribution of the Andreev reflection eigenvalues, which we then apply to electrical conduction in the next section. For each symmetry class we first determine the polar decomposition of $r$ in terms of the $R_{n}$'s. The resulting invariant measure $d\mu(r)\propto P(\{R_{n}\})\prod_{n}dR_{n}$ then gives the probability distribution $P(\{R_{n}\})$ of the Andreev reflection eigenvalues in the corresponding circular ensemble.

\subsection{Class D, ensemble CRE}
\label{CRE}

In the absence of time-reversal and spin-rotation symmetry the scattering matrix has the polar decomposition
\begin{equation}
r=\begin{pmatrix}
U&0\\
0&U^{\ast}
\end{pmatrix}
\begin{pmatrix}
\Gamma&-i\Lambda\\
i\Lambda&\Gamma
\end{pmatrix}
\begin{pmatrix}
V^{\dagger}&0\\
0&V^{\rm T}
\end{pmatrix}.\label{polarD}
\end{equation}
The $N\times N$ matrices $U,V$ are unitary and the $N\times N$ matrices $\Lambda$ and $\Gamma$ are real, to satisfy the particle-hole symmetry condition \eqref{ehsymmetry}. Unitarity of $r$ requires, in addition to $\Gamma^{\rm T}\Gamma+\Lambda^{\rm T}\Lambda=1$, that $\Lambda^{\rm T}\Gamma=-\Gamma^{\rm T}\Lambda$ is antisymmetric. As derived in Ref.\ \onlinecite{Ber09a}, the matrices $\Lambda$ and $\Gamma$ must therefore have a $2\times 2$ block diagonal structure.

For $N=2M$ even and $Q=1$ one has $\Lambda=\Lambda_{M}$, $\Gamma=\Gamma_{M}$ with
\begin{align}
&\Lambda_{M}=\bigoplus_{n=1}^{M}
\begin{pmatrix}
\sin\alpha_{n}&0\\
0&\sin\alpha_{n}
\end{pmatrix}=\bigoplus_{n=1}^{M}\sigma_{0}\sin\alpha_{n},\label{LambdaMdef}\\
&\Gamma_{M}=\bigoplus_{n=1}^{M}
\begin{pmatrix}
0&\cos\alpha_{n}\\
-\cos\alpha_{n}&0
\end{pmatrix}=\bigoplus_{n=1}^{M}i\sigma_{y}\cos\alpha_{n}.\label{GammaMdef}
\end{align}
The $2\times 2$ Pauli matrices are $\sigma_{x},\sigma_{y},\sigma_{z}$ (with $\sigma_{0}$ the $2\times 2$ unit matrix). The real angles $\alpha_{n}\in(0,2\pi)$ determine the Andreev reflection eigenvalues $R_{n}=\sin^{2}\alpha_{n}$. These are all twofold degenerate.

The parameterization derived in Ref.\ \onlinecite{Ber09a} has $\Lambda_{M}\propto i\sigma_{y}$ and $\Gamma_{M}\propto\sigma_{0}$. The present, equivalent, form is chosen because it is more easily extended to symmetry class DIII (where an additional symmetry condition applies). For the same reason, we parameterize the diagonal entries in terms of the angles $\alpha_{n}$ rather than in terms of $\sqrt{R_{n}}$ and $\sqrt{1-R_{n}}$. (The sign of the terms $\sin\alpha_{n}$, $\cos\alpha_{n}$ cannot be fixed in class DIII.)

To check that the polar decomposition \eqref{polarD}--\eqref{GammaMdef} indeed gives ${\rm Det}\,r=1$, one can use the identities ${\rm Det}\,AB=({\rm Det}\,A)({\rm Det}\,B)$ and
\begin{equation}
{\rm Det}\,\begin{pmatrix}
A&B\\
C&D
\end{pmatrix}={\rm Det}\,(AD-ACA^{-1}B).\label{foldingDet}
\end{equation}

For $N=2M$ even and $Q=-1$ one has
\begin{equation}
\Lambda=\Lambda_{M-1}\oplus\begin{pmatrix}
0&0\\
0&1
\end{pmatrix},\;\;
\Gamma=\Gamma_{M-1}\oplus\begin{pmatrix}
1&0\\
0&0
\end{pmatrix},\label{N2MQmin1}
\end{equation}
so in addition to $M-1$ two-fold degenerate eigenvalues $R_{1},R_{2},\ldots R_{M-1}$ there is one nondegenerate eigenvalue equal to $0$ and one nondegenerate eigenvalue equal to $1$. It is this unit Andreev reflection eigenvalue which contributes a factor $-1$ to ${\rm Det}\,r$.

For $N=2M+1$ odd there are $M$ two-fold degenerate eigenvalues $R_{1},R_{2},\ldots R_{M}$ plus one nondegenerate eigenvalue equal to $q=(1-Q)/2$,
\begin{equation}
\Lambda=\Lambda_{M}\oplus (q),\;\;
\Gamma=\Gamma_{M}\oplus (1-q).\label{N2Mmin1}
\end{equation}
The nondegenerate eigenvalue equals $1$ in the topologically nontrivial phase and $0$ otherwise. Again, it is the unit Andreev reflection eigenvalue which gives ${\rm Det}\,r=-1$.

The calculation of the invariant measure from these polar decompositions is outlined in App.\ \ref{AppA}. The resulting probability distributions of the twofold degenerate Andreev reflection eigenvalues in the CRE are
\begin{align}
P(\{R_{n}\})\propto&\prod_{i<j=1}^{M}(R_{i}-R_{j})^{4},\nonumber\\
&{\rm if}\;\; N=2M\;\;{\rm and}\;\; Q=1,\label{PCREa}\\
P(\{R_{n}\})\propto&\prod_{i<j=1}^{M-1}(R_{i}-R_{j})^{4}\prod_{k=1}^{M-1}R_{k}^{2}(1-R_{k})^{2},\nonumber\\
&{\rm if}\;\; N=2M\;\;{\rm and}\;\; Q=-1,\label{PCREb}\\
P(\{R_{n}\})\propto&\prod_{i<j=1}^{M}(R_{i}-R_{j})^{4}\prod_{k=1}^{M}[R_{k}-\tfrac{1}{2}(1-Q)]^{2},\nonumber\\
&{\rm if}\;\; N=2M+1.\label{PCREc}
\end{align}
The degenerate Andreev reflection eigenvalues repel each other with the fourth power of their separation. In addition there is a repulsion with the second power of the separation to each of the nondegenerate eigenvalues, pinned at $0$ or $1$.

\subsection{Class DIII, ensemble T-CRE}
\label{TCRE}

In the presence of time-reversal symmetry the scattering matrix should also satisfy the condition \eqref{rSigmayrT}, which implies that $i\Sigma_{y}r$ is antisymmetric. The polar decomposition which respects this symmetry condition (as well as the condition \eqref{ehsymmetry} for particle-hole symmetry) is
\begin{equation}
i\Sigma_{y}r=\begin{pmatrix}
\Omega&0\\
0&\Omega^{\ast}
\end{pmatrix}
\begin{pmatrix}
\Gamma&-i\Lambda\\
i\Lambda&\Gamma\
\end{pmatrix}
\begin{pmatrix}
\Omega^{\rm T}&0\\
0&\Omega^{\dagger}
\end{pmatrix},\label{iSigmayr}
\end{equation}
with $\Omega$ an $N\times N$ unitary matrix. Unitarity still requires that $\Lambda^{\rm T}\Gamma$ is antisymmetric, while time-reversal symmetry requires $\Gamma^{\rm T}=-\Gamma$, $\Lambda^{\rm T}=\Lambda$.

The number of channels $N=2M$ is even, with $M$ the number of channels per spin. Each reflection eigenvalue has a two-fold degeneracy, including those equal to $0$ or $1$. This Kramers degeneracy due to time-reversal symmetry\cite{Bar08} in class DIII \textit{replaces} the B\'{e}ri degeneracy due to electron-hole symmetry\cite{Ber09a} in class D --- it is not an additional degeneracy. The topological quantum number\cite{Ful11} $Q={\rm Pf}\,i\Sigma_{y}r$ can be calculated using the identity
\begin{equation}
{\rm Pf}\,\begin{pmatrix}
0&a&b&c\\
-a&0&d&e\\
-b&-d&0&f\\
-c&-e&-f&0
\end{pmatrix}=af-be+cd,\label{Pfidentity}
\end{equation}
for scalars $a,b,c,d,e,f$, and also the formulas ${\rm Pf}\,XYX^{\rm T}=({\rm Det}\,X)({\rm Pf}\,Y)$, ${\rm Pf}\,\bigoplus_{n}Y_{n}=\prod_{n}{\rm Pf}\,Y_{n}$ (valid for arbitrary square matrices $X$ and antisymmetric matrices $Y,Y_{n}$). 

For $Q=1$ we take $\Lambda=\Lambda_{M}$ and $\Gamma=\Gamma_{M}$ from Eqs.\ \eqref{LambdaMdef} and \eqref{GammaMdef}. The Pfaffian of the matrix \eqref{iSigmayr} is always $+1$, so this polar decomposition describes the topologically trivial phase. In order to reach the topologically nontrivial phase, we include a twofold degenerate eigenvalue equal to unity, but with a $\sigma_{z}$ matrix rather than a $\sigma_{0}$ matrix: $\Lambda=\Lambda_{M-1}\oplus{\rm diag}\,(1,-1)$, $\Gamma=\Gamma_{M-1}\oplus{\rm diag}\,(0,0)$. Then the Pfaffian is $-1$.

As derived in App.\ \ref{AppA}, the distribution of the $M$ degenerate Andreev reflection eigenvalues in the T-CRE is given most compactly in terms of the variables $\xi_{n}=\sin\alpha_{n}\in(-1,1)$ (with $R_{n}=\xi_{n}^{2}$). For $Q=1$ the result is
\begin{equation}
P(\{\xi_{n}\})\propto\prod_{i<j=1}^{M}(\xi_{i}-\xi_{j})^{4}.\label{PTCREa}
\end{equation}
Notice that there is no repulsion of pairs of Andreev reflection eigenvalues: If $\xi_{i}\rightarrow -\xi_{j}$ then $R_{i}\rightarrow R_{j}$ and yet the probability distribution does not vanish.

For $Q=-1$ one pair of eigenvalues is pinned at $R_{M}=1\Rightarrow\xi_{M}=1$. The distribution of the remaining $M-1$ degenerate eigenvalues is
\begin{equation}
P(\{\xi_{n}\})\propto\prod_{i<j=1}^{M-1}(\xi_{i}-\xi_{j})^{4}\prod_{k=1}^{M-1}(1-\xi_{k}^{2})^{2}.\label{PTCREb}
\end{equation}
While pairs of Andreev reflection eigenvalues $R_{n}\in(0,1)$ in the T-CRE do not repel each other, they are repelled from the eigenvalue pinned at $R_{M}=1$, with the same second power of the separation as in the CRE. 

\subsection{Class C, ensemble CQE}
\label{CQE}

For completeness we also consider the two symmetry classes C and CI without a topological phase. Then spin-rotation symmetry is preserved, so it is sufficient to consider a single spin degree of freedom, say an electron in the spin-up band coupled to a hole in the spin-down band. The reflection matrix for this scattering process has dimension $2M\times 2M$, where $M$ only counts the orbital degree of freedom. Each reflection eigenvalue has a twofold spin degeneracy.

The polar decomposition of the reflection matrix reads
\begin{equation}
r=\begin{pmatrix}
U&0\\
0&U^{\ast}
\end{pmatrix}
\begin{pmatrix}
\cos\bm{\alpha}&i\sin\bm{\alpha}\\
i\sin\bm{\alpha}&\cos\bm{\alpha}
\end{pmatrix}
\begin{pmatrix}
V^{\dagger}&0\\
0&V^{\rm T}
\end{pmatrix},\label{polarC}
\end{equation}
as required by unitarity and the particle-hole symmetry condition \eqref{ehsymmetry2}. Here $U,V$ are unitary $M\times M$ matrices and $\bm{\alpha}={\rm diag}\,(\alpha_{1},\alpha_{2},\ldots\alpha_{M})$ is the diagonal matrix of real angles that determine the reflection eigenvalues $R_{n}=\sin^{2}\alpha_{n}$. As before, we might have replaced $\sin\alpha_{n}\mapsto\sqrt{R_{n}}$ and $\cos\alpha_{n}\mapsto\sqrt{1-R_{n}}$ in this polar decomposition for class C, but not when we additionally impose time-reversal symmetry (in class CI).

The factor $i$ in Eq.\ \eqref{polarC} accounts for the $\pi/2$ phase shift associated with Andreev reflection of an electron into a hole from the opposite spin band. No such factor appears in the polar decomposition \eqref{polarD} in the absence of spin-rotation symmetry, because there it can be absorbed in the unitary matrices (which in that case contain both spin bands for electrons and holes).

The probability distribution of the Andreev reflection eigenvalues in the CQE was calculated previously by Khaymovich et al.\cite{Kha10} We find
\begin{equation}
P(\{R_{n}\})\propto\prod_{i<j=1}^{M}|R_{i}-R_{j}|,\label{PCQE}
\end{equation}
in agreement with Ref.\ \onlinecite{Kha10} (up to an evident misprint, $\prod_{i\neq j}$ instead of $\prod_{i<j}$).

\subsection{Class CI, ensemble T-CQE}
\label{TCQE}

Finally, in class CI we have the additional requirement \eqref{TRSsymmetry} of time-reversal symmetry. The polar decomposition becomes
\begin{equation}
r=\begin{pmatrix}
U&0\\
0&U^{\ast}
\end{pmatrix}
\begin{pmatrix}
\cos\bm{\alpha}&i\sin\bm{\alpha}\\
i\sin\bm{\alpha}&\cos\bm{\alpha}
\end{pmatrix}
\begin{pmatrix}
U^{\rm T}&0\\
0&U^{\dagger}
\end{pmatrix}.\label{polarCI}
\end{equation}

The distribution of the $R_{n}$'s in the T-CQE (each doubly degenerate) is again given most compactly in terms of the variables $\xi_{n}=\sin\alpha_{n}\in(-1,1)$ (with $R_{n}=\xi_{n}^{2}$). We find
\begin{equation}
P(\{\xi_{n}\})\propto\prod_{i<j=1}^{M}|\xi_{i}-\xi_{j}|.\label{PTCQE}
\end{equation}
As in the T-CRE, there is no repulsion between pairs of Andreev reflection eigenvalues in the presence of time-reversal symmetry.

\section{Dependence on topological quantum number of the conductance distribution}
\label{conductance}

\subsection{Broken time-reversal symmetry}
\label{brokenTRS}

\begin{figure}[tb]
\centerline{\includegraphics[width=0.8\linewidth]{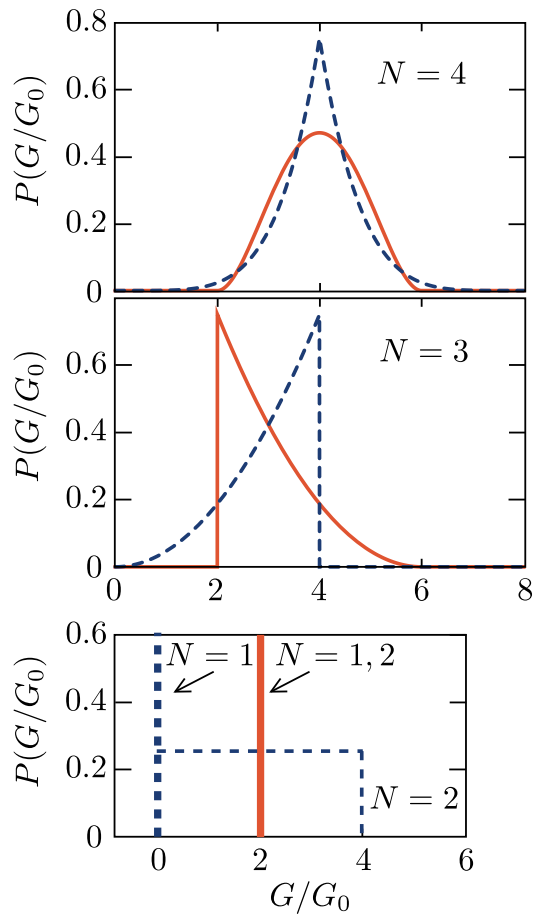}}
\caption{\label{fig_Gplot}
Probability distribution of the conductance in the CRE, for channel numbers $N=1,2,3,4$ and topological charges $Q=-1$ (red solid curves) and $Q=+1$ (blue dashed curves). In the lower panel the thick vertical lines indicate a delta-function distribution.
}
\end{figure}

From the probability distributions $P(\{R_{n}\})$ in Secs.\ \ref{CRE} and \ref{TCRE} we readily calculate the distribution $P(G)$ of the conductance \eqref{BTK}, in both the topologically trivial and nontrivial phases. We first consider the case without time-reversal symmetry (class D, ensemble CRE). Using Eqs.\ \eqref{PCREa}--\eqref{PCREc} we obtain the distributions for the first few channel numbers $N=1,2,3,4$. The results are plotted in Fig.\ \ref{fig_Gplot}, and given by: 
\begin{itemize}
\item For $N=1$, the conductance $G/G_{0}=1-Q$ without statistical fluctuations.\cite{Ber09a,Ber09b}
\item For $N=2$, the conductance $G/G_{0}=2$ for $Q=-1$ without statistical fluctuations; if $Q=1$, instead $G/G_{0}=4g$ with $P(g)=1$.
\item For $N=3$ the conductance $G/G_{0}=1-Q+4g$, with $P(g)=3(\frac{1}{2}-\frac{1}{2}Q-g)^{2}$.
\item For $N=4$ the conductance $G/G_{0}=2+4g$ if $Q=-1$, with $P(g)=30\, g^{2}(1-g)^{2}$, while if $Q=1$ we have $G/G_{0}=8g$ with $P(g)=6(1-|1-2 g|)^5$.
\end{itemize}
In these expressions, $g$ denotes a stochastic variable in the range $(0,1)$.

\begin{table}
\centering
\begin{tabular}{ | c || c | c | c | c | c|}
\hline
&\multicolumn{5}{|c|}{$\langle\!\langle(G/G_{0})^{p}\rangle\!\rangle$} \\
& $p=1$ &  $p=2$ & $p=3$ & $p=4$ & $p=5$ \\ \hline
$N=1$ & $\bm{0\;\vdots\;2}$ & $0\;\vdots\;0$ & $0\;\vdots\;0$ & $0\;\vdots\;0$ & $0\;\vdots\;0$ \\ \hline
$N=2$ & $2\;\vdots\;2$ & $\bm{\frac{4}{3}\;\vdots\;0}$ & $0\;\vdots\;0$ & $\mathord{-}\frac{32}{15}\;\vdots\;0\hphantom{-\;\;}$ & $0\;\vdots\;0$ \\[2pt] \hline
$N=3$ & $3\;\vdots\;3$ & $\frac{3}{5}\;\vdots\;\frac{3}{5}$ & $\bm{\mathord{-}\frac{2}{5}\;\vdots\;\frac{2}{5}}\hphantom{\mathord{-}}$ & $\frac{6}{175}\;\vdots\;\frac{6}{175}$ & $\hphantom{\mathord{-}}\frac{24}{35}\;\vdots\;\mathord{-}\frac{24}{35}$ \\[2pt] \hline
$N=4$ & $4\;\vdots\;4$ & $\frac{4}{7}\;\vdots\;\frac{4}{7}$ & $0\;\vdots\;0$ & $\hphantom{-}\bm{\frac{176}{735}\;\vdots\;\mathord{-}\frac{32}{147}}$ & $0\;\vdots\;0$ \\[2pt] \hline
$N=5$ & $5\;\vdots\;5$ & $\frac{5}{9}\;\vdots\;\frac{5}{9}$ & $0\;\vdots\;0$ & $\frac{10}{2079}\;\vdots\;\frac{10}{2079}$  & $\bm{\mathord{-}\frac{8}{63}\;\vdots\;\frac{8}{63}}\hphantom{\mathord{-}}$ \\[2pt] \hline
\end{tabular}
\caption{First five cumulants ($p\leq 5$) of conductance in the CRE, calculated for number of modes $N\leq 5$ and topological quantum number $Q$. (The first entry in each cell is for $Q=1$, the second entry is for $Q=-1$.) The conductance distribution depends on $Q$ starting from the $N$-th cumulant (bold).}
\label{table_CRE}
\end{table}

From Fig.\ \ref{fig_Gplot} we see that upon increasing $N$, the conductance distributions for $Q=1$ and $Q=-1$ become more and more similar. To quantify the difference, we list in Table \ref{table_CRE} the first few cumulants $\langle\!\langle G^{p}\rangle\!\rangle$ of $P(G)$ for several values of $N$. Inspection of the table brings us to propose that:
\begin{itemize}
\item[]
\textit{The cumulant of order $p$ of the $N$-mode conductance in the CRE is independent of the topological charge for $p<N$.}
\end{itemize}
A proof for arbitrary $N$ is given in App.\ \ref{AppB}.

\begin{figure}[tb]
\centerline{\includegraphics[width=0.8\linewidth]{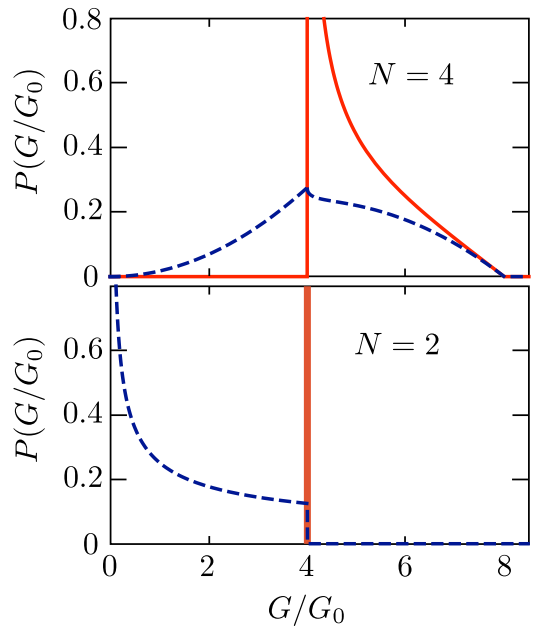}}
\caption{\label{fig_Gplotb}
Same as Fig.\ \ref{fig_Gplot}, for the T-CRE.
}
\end{figure}

\subsection{Preserved time-reversal symmetry}
\label{preservedTRS}

In the presence of time-reversal symmetry (class DIII, ensemble T-CRE) we can similarly calculate the conductance distribution from Eqs.\ \eqref{PTCREa} and \eqref{PTCREb}. For small $N$ we find:
\begin{itemize}
\item For $N=2$, the conductance $G/G_{0}=4$ without statistical fluctuations if $Q=-1$, while if $Q=1$ we have $G/G_{0}=4g$ with $P(g)=\frac{1}{2}g^{-1/2}$.
\item For $N=4$ the conductance $G/G_{0}=4+4g$ if $Q=-1$, with $P(g)=\frac{15}{16}(1-g)^{2}g^{-1/2}$; if $Q=1$, instead $G/G_{0}=8g$ with $P(g)$ plotted in Fig.\ \ref{fig_Gplotb}, upper panel. (The analytic expression is lengthy.)
\end{itemize}
The analogous theorem for the $Q$-independence of low-order cumulants now reads:
\begin{itemize}
\item[]
\textit{The cumulant of order $p$ of the $N$-mode conductance in the T-CRE is independent of the topological charge for $p<N/2$.}
\end{itemize}
A proof for arbitrary (even) $N$ is given also in App.\ \ref{AppB}.

\subsection{Weak localization and UCF}
\label{UCF}

Weak localization and universal conductance fluctuations (UCF) refer to the average and to the variance of the conductance in the large-$N$ limit. Since the dependence on the topological charge is nonpertubative in $N$, these two effects cannot contain any information on whether the superconductor is in a topological phase or not. As a check, we have calculated the average $\langle G\rangle$ and the variance ${\rm Var}\,G=\langle G^{2}\rangle-\langle G\rangle^{2}$ for $N\gg 1$, directly from the probability distribution of the Andreev reflection eigenvalues. This calculation also allows us to verify a conjecture from Ref.\ \onlinecite{Alt97} on the UCF in the presence of time-reversal symmetry. Since the calculation follows established methods in random-matrix theory,\cite{Bee97} we only give the results. 

The weak-localization correction $\delta G=G-NG_{0}$ to the conductance vanishes in the CRE and CQE, while $\delta G/G_{0}=\frac{1}{2},-1$ in the T-CRE and T-CQE, respectively. The UCF are given by ${\rm Var}\,G/G_{0}=\frac{1}{2},1,2,4$ in the CRE, T-CRE, CQE, and T-CQE, respectively. These $Q$-independent results are in full agreement with Ref.\ \onlinecite{Alt97}.

All these results assume that the proximity to the superconductor does not induce an excitation gap in the quantum dot. In the CRE and CQE this is realized by the pair-breaking magnetic field. In the T-CRE and T-CQE we need a $\pi$-junction to close the gap: Two NS interfaces, coupled equally well to the quantum dot and with a $\pi$ phase difference of the superconducting phase.\cite{Alt97} For a single NS interface in zero magnetic field, the presence of an excitation gap does not change the value of $\delta G$, but the variance of the conductance is changed into\cite{Bee93} ${\rm Var}\,G/G_{0}=9/4\beta$, with $\beta=1$ or $\beta=4$ in the presence or absence of spin-rotation symmetry. Notice that time-reversal symmetry breaking then has only a relatively small 10\% effect on the UCF,\cite{Bee97} while in the absence of the excitation gap the effect on the variance is a factor of two.\cite{Alt97}

\section{Conclusion and comparison with a model Hamiltonian}
\label{conclude}

In conclusion, we have shown that the distribution $P(G)$ of the electrical conductance in a quantum dot connecting a normal-metal to a superconducting electrode has a striking dependence on the topological quantum number $Q$ of the superconductor, but only if the number of modes $N$ in which the current is injected is sufficiently small. In the absence of time-reversal and spin-rotation symmetry, the distributions for $Q=-1$ and $Q=+1$ differ in the average conductance for $N=1$, in the variance for $N=2$, in the skewness for $N=3$, and in the kurtosis for $N=4$. More generally, the dependence appears in the cumulant of order $N$ or $N/2$, depending on whether time-reversal symmetry is broken or not.

\begin{figure}[tb]
\centerline{\includegraphics[width=0.9\linewidth]{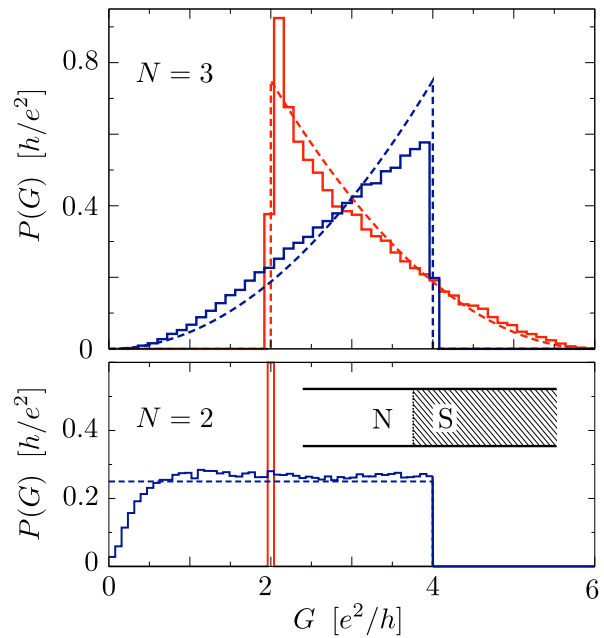}}
\caption{\label{fig_histograms}
Comparison of the probability distribution of the electrical conductance as predicted by RMT (dashed curves) and as resulting from numerical simulation of the model Hamiltonian \eqref{HBdG} (solid histograms). The simulation is for the disordered normal-metal--superconductor junction shown in the inset. The number of propagating modes in the normal region is $N=2$ (lower panel) and $N=3$ (upper panel), while the red and blue curves are for topological quantum number $Q=-1$ and $Q=+1$, respectively. The disorder strength is fixed at $U_{0}=130\,E_{\rm so}$ for $N=2$ and $U_{0}=100\,E_{\rm so}$ for $N=3$. The values used for Fermi energy and Zeeman energy (in units of $E_{\rm so}$) are as follows. For $N=2$: $E_{F}=12$, $E_{Z}=3.8$ ($Q=1$) and
$E_{F}=13$,
$E_{Z}=9$ ($Q=-1$). For $N=3$:
$E_{F}=19$,
$E_{Z}=3.8$ ($Q=1$)
and $E_{F}=19$,
$E_{Z}=8$ ($Q=-1$).
}
\end{figure}

The system we have considered (Fig.\ \ref{fig_QDNS}) is constructed to ensure chaotic scattering, which is the requirement for application of the circular ensembles of RMT. Systems of present experimental focus in the search for Majorana bound states have a simpler wire geometry, without the quantum dot (Fig.\ \ref{fig_histograms}, inset). Impurity scattering within a superconducting coherence length from the NS interface can still lead to chaotic dynamics, at least if the number of modes is sufficiently small that they are fully mixed by the disorder.

To test the applicability of our RMT results to such a system we have performed numerical simulations of the model Hamiltonian of Refs.\ \onlinecite{Lut10a,Ore10}, which describes an InAs nanowire on an Al or Nb substrate. The Bogoliubov-De Gennes Hamiltonian
\begin{align}
{\cal H}&=\begin{pmatrix}
1&0\\
0&\sigma_{y}
\end{pmatrix}
\begin{pmatrix}
H_{\rm R}-E_{F}&\Delta\\
\Delta^{\ast}&E_{F}-\sigma_{y}H_{\rm R}^{\ast}\sigma_{y}
\end{pmatrix}
\begin{pmatrix}
1&0\\
0&\sigma_{y}
\end{pmatrix}\nonumber\\
&=\begin{pmatrix}
H_{\rm R}-E_{F}&\Delta\sigma_{y}\\
\Delta^{\ast}\sigma_{y}&E_{F}-H_{\rm R}^{\ast}
\end{pmatrix}
\label{HBdG}
\end{align}
couples electron and hole excitations near the Fermi energy $E_{F}$ through an \textit{s}-wave superconducting order parameter $\Delta$. (We have made a unitary transformation to ensure that the condition for particle-hole symmetry has the form used in the preceding sections.)

The excitations are confined to a wire of width $W$ in the $x-y$ plane of the semiconductor surface inversion layer, where their dynamics is governed by the Rashba Hamiltonian
\begin{equation}
H_{\rm R}=\frac{\bm{p}^{2}}{2m_{\rm eff}}+U(\bm{r})+\frac{\alpha_{\rm so}}{\hbar}(\sigma_{x}p_{y}-\sigma_{y}p_{x})+\tfrac{1}{2}g_{\rm eff}\mu_{B}B\sigma_{x}.\label{HRashba}
\end{equation}
The spin is coupled to the momentum $\bm{p}=-i\hbar\partial/\partial{\bm r}$ by the Rashba effect, and polarized through the Zeeman effect by a magnetic field $B$ parallel to the wire (in the $x$-direction). Characteristic length and energy scales are $l_{\rm so}=\hbar^{2}/m_{\rm eff}\alpha_{\rm so}$ and $E_{\rm so}=m_{\rm eff}\alpha_{\rm so}^{2}/\hbar^{2}$. Typical values in InAs are $l_{\rm so}=100\,{\rm nm}$, $E_{\rm so}=0.1\,{\rm meV}$, $E_{Z}=\frac{1}{2}g_{\rm eff}\mu_{B}=1\,{\rm meV}$ at $B=1\,{\rm T}$.

We have solved the scattering problem numerically\cite{Wim09} by discretizing the Hamiltonian \eqref{HBdG} on a square lattice (lattice constant $a=l_{\rm so}/20$), with an electrostatic disorder potential $U(x,y)$ that varies randomly from site to site, distributed uniformly in the interval $(-U_{0},U_{0})$. The disordered superconducting wire (width $W=20\,a$, length $L=800\,a$, $\Delta=4\,E_{\rm so}$) is connected at two ends to ideal normal-metal leads, obtained by setting $\Delta,U_{0}\equiv 0$ for $x<0$, $x>L$. The length $L$ was chosen large enough that the transmission probability through the wire was $< 10^{-2}$.

Results for the probability distribution of the electrical conductance are shown in Fig.\ \ref{fig_histograms}, for $N=2,3$ and $Q=-1,1$. (For $N=1$ we simply find the two delta-function distributions at $G=(e^{2}/h)(1-Q)$, as expected.) The histograms were obtained by averaging over $10^{5}$ disorder realizations, conditionally on the value of the topological quantum number $Q=\pm 1$ (calculated from $Q={\rm sign}\,{\rm Det}\,r$, as in Ref.\ \onlinecite{Akh10}.) The agreement with the predictions from RMT is quite satisfactory.

\acknowledgments

This research was supported by the Dutch Science Foundation NWO/FOM, by the Deutscher Akademischer Austausch Dienst DAAD, and by an ERC Advanced Investigator Grant.
\newpage

\appendix

\section{Calculation of the invariant measure}
\label{AppA}

In this appendix we derive the probability distributions of the Andreev reflection eigenvalues in the circular ensembles, given in Sec.\ \ref{RMT}. We work out the derivation for the symmetry classes D and DIII, for an even number $N=2M$ of modes and for topological charge $Q=1$, following established methods of random-matrix theory.\cite{For10} The calculations for the other ensembles are entirely analogous, so we do not present them here.

The circular ensembles are characterized by a uniform probability distribution, constrained only by fundamental symmetries. Uniformity in the unitary group is defined with respect to the invariant (Haar) measure $d\mu(r)=r^{\dagger}dr\equiv \delta r$. Since the polar decompositions in Sec.\ \ref{RMT} give a parametrization of the (unitary) reflection matrix $r$ in terms of the angles $\alpha_n$, we can transform the measure into $d\mu( r)=J \prod_i dp_i \prod_n d\alpha_n$. The $p_i$'s are the degrees of freedom of the matrices of eigenvectors and $J$ is the Jacobian of the transformation. From this expression the probability distribution of the angles $\alpha_n$ follows upon integration over the $p_i$'s,
\begin{equation}
P(\{\alpha_n\})\propto \int J \prod_i dp_i,
\label{Adistribution}
\end{equation} 
up to a normalization constant. 

The Jacobian can be found from the metric tensor $g_{\mu\nu}$, which can be read off from the trace $\text{Tr}\,\delta r \delta r^\dag$ when it is expressed in terms of the infinitesimals $d\alpha_n$ and $dp_i$ (collectively denoted as $dx_{\mu}$):
\begin{equation}
\text{Tr}\,\delta r \delta r^\dag=\sum_{\mu,\nu}g_{\mu\nu}dx_{\mu}dx_{\nu},\;\;J=|{\rm Det}\,g_{\mu\nu}|^{1/2}.\label{metric}
\end{equation}
We carry out this calculation first for class D and then for class DIII.

\subsection{Class D (ensemble CRE)}
\label{appA_D}

In view of the polar decomposition \eqref{polarD} one has
\begin{align}
\begin{pmatrix}
U^{\dag}&0\\
0&U^{\rm T}
\end{pmatrix}
dr
\begin{pmatrix}
V&0\\
0&V^{\ast}
\end{pmatrix}
={}&
\begin{pmatrix}
\delta U&0\\
0&\delta U^*
\end{pmatrix}
L+dL\nonumber\\
&-L
\begin{pmatrix}
\delta V&0\\
0&\delta V^*
\end{pmatrix},\label{eqdL}
\end{align}
where we abbreviated
\begin{equation}
L=\begin{pmatrix}
\Gamma &-i \Lambda \\
i\Lambda &\Gamma
\end{pmatrix}.
\end{equation}
The quantities $\delta U=U^\dag dU$ and $\delta V=V^\dag dV$ represent measures on the eigenvector manifolds. We used that $d(V^{\dagger}V)=0\Rightarrow (dV^{\dagger})V=(\delta V)^{\dagger}=-\delta V$. 

Substitution of Eq.\ \eqref{eqdL} into $\text{Tr}\,\delta r \delta r^\dag=\text{Tr}\,dr dr^\dag$ gives
\begin{align}
\text{Tr}\,\delta r \delta r^\dag={}&2\,\text{Tr}\,L 
\begin{pmatrix}
\delta V&0\\
0&\delta V^*
\end{pmatrix}
L^\dag
\begin{pmatrix}
\delta U&0\\
0&\delta U^*
\end{pmatrix}
\nonumber\\
& -\text{Tr}\,[\delta U^2 + (\delta U^*)^2+\delta V^2+(\delta V^*)^2]\nonumber\\
&+\text{Tr}\,dLdL^\dag.
\end{align}
(All other cross terms vanish.) In terms of $\Gamma$ and $\Lambda$ this can be expressed as a sum of five traces,
\begin{align}
\text{Tr}\,\delta r \delta r^\dag={}&
\text{Tr}\,(\Gamma \delta V -\delta U \Gamma ) (\Gamma^{\rm T} \delta U -\delta V \Gamma^{\rm T})\nonumber\\
&+{\rm Tr}\,(\Gamma \delta V^* -\delta U^* \Gamma ) (\Gamma^{\rm T} \delta U^* -\delta V^* \Gamma^{\rm T})\nonumber\\
&+{\rm Tr}\,(\Lambda \delta V^* -\delta U \Lambda ) (\Lambda^{\rm T} \delta U -\delta V^* \Lambda^{\rm T})\nonumber\\
&+{\rm Tr}\,(\Lambda \delta V -\delta U^* \Lambda ) (\Lambda^{\rm T} \delta U^* -\delta V \Lambda^{\rm T})\nonumber\\
&+\text{Tr}\,dLdL^{\rm T}\nonumber\\
\equiv{}&T_1+T_2+T_3+T_4+T_5.\label{eqtraces}
\end{align}

Each of the traces in Eq.\ \eqref{eqtraces} is of the form $\text{Tr}\,AA^\dag=\sum_{ij} |A_{ij}|^2$, and is therefore real. Since the second line is the complex conjugate of the first line and the fourth line is the complex conjugate of the third line, their traces are the same, hence $T_1=T_2$ and $T_3=T_4$. For the evaluation of the expression we need to distinguish between the different values of the topological quantum number and between  the cases of odd and even number of channels. 

We work out the calculation for $N=2M$ even and $Q=1$, when $\Lambda$ and $\Gamma$ are given by Eqs. \eqref{LambdaMdef} and \eqref{GammaMdef}. The trace $T_5$ is easiest to evaluate,
\begin{align}
T_5&=\sum_{i,j} |dL_{ij}|^2=4 \sum_{i=1}^{M} d\alpha_i^2.\label{T5result}
\end{align}
This trace contributes a diagonal block to the metric tensor and a constant factor to the Jacobian, for $M$ independent real parameters. The other two traces $T_1$ and $T_3$ require more work,
\begin{widetext}
\begin{align}
T_1={}&\sum_{r<s=1}^M\sum_{k,l=0}^1 \left\{ \tfrac{1}{2}(\cos \alpha_{r}+\cos \alpha_s)^2  |\delta U_{2r-k,2s-l}-(-1)^{k+l}\delta V_{2r-\bar{k},2s-\bar{l}}|^2\right. \nonumber\\
&+\left.\tfrac{1}{2}(\cos \alpha_{r}-\cos \alpha_s)^2  |\delta U_{2r-k,2s-l}+(-1)^{k+l}\delta V_{2r-\bar{k},2s-\bar{l}}|^2\right\}\nonumber\\
&+\sum_m^M\sum_{k,l=0}^1 \cos^2\alpha_m |\delta V_{2m-k,2m-l}-(-1)^{k+l}\delta U_{2m-\bar{k},2m-\bar{l}}|^2,\\
T_3={}&\sum_{r<s=1}^M \sum_{k,l=0}^1 \left\{ \tfrac{1}{2}(\sin \alpha_{r}+\sin \alpha_s)^2  |\delta V^*_{2r-k,2s-l}-\delta U_{2r-k,2s-l}|^2\right.\nonumber\\
&+\left. \tfrac{1}{2}(\sin \alpha_{r}-\sin \alpha_s)^2  |\delta V^*_{2r-k,2s-l}+\delta U_{2r-k,2s-l}|^2\right\}\nonumber\\
&+\sum_m^M  \sum_{k,l=0}^1 \sin^2\alpha_m |\delta V^*_{2m-k,2m-l}-\delta U_{2m-k,2m-l}|^2\label{Tr3}.
\end{align}
\end{widetext}
We denote $\bar{k}=0,1$ for $k=1,0$. 

We group the indices of the matrices $\delta U$ and $\delta V$ into $2\times 2$ blocks, and consider first the off-diagonal blocks. For these we can choose as independent parameters
\begin{align}
&\delta U_{2r,2s},\;\;\delta U_{2r-1,2s},\;\;\delta U_{2r,2s-1},\;\;\delta U_{2r-1,2s-1},\nonumber\\
&\delta V_{2r,2s},\;\;\delta V_{2r-1,2s},\;\;\delta V_{2r,2s-1},\;\;\delta V_{2r-1,2s-1},\nonumber
\end{align}
with $1\le r<s\le M$. The real and imaginary parts, denoted by $\delta U^{\rm R},\delta U^{\rm I},\delta V^{\rm R},\delta V^{\rm I}$, produce a total of $8M(M-1)$ independent parameters. The contribution to $\text{Tr}\,\delta r \delta r^\dag$ for given values of $r$ and $s$ has the form
\begin{align}
&\sum_{k=2r-1}^{2r}\sum_{l=2s-1}^{2s}\biggl\{4\left[(\delta U^{\rm R}_{kl})^2+(\delta U^{\rm I}_{kl})^2+(\delta V^{\rm R}_{kl})^2+(\delta V^{\rm I}_{kl})^2\right]\nonumber\\
&\quad+2a \left[ \delta V_{kl}^{\rm R}\delta U_{kl}^{\rm R}-\delta V_{kl}^{\rm I}\delta U_{kl}^{\rm I}\right]\biggr\}\nonumber\\
&+2b\left[ \delta V_{2r-1,2s-1}^{\rm R}\delta U_{2r,2s}^{\rm R}+\delta V_{2r-1,2s-1}^{\rm I}\delta U_{2r,2s}^{\rm I}\right]\nonumber\\
&+2b\left[ \delta V_{2r,2s}^{\rm R}\delta U_{2r-1,2s-1}^{\rm R}+\delta V_{2r,2s}^{\rm I}\delta U_{2r-1,2s-1}^{\rm I}\right]\nonumber\\
&-2b\left[ \delta V_{2r-1,2s}^{\rm R}\delta U_{2r,2s-1}^{\rm R}+\delta V_{2r-1,2s}^{\rm I}\delta U_{2r,2s-1}^{\rm I}\right]\nonumber\\
&-2b\left[ \delta V_{2r,2s-1}^{\rm R}\delta U_{2r-1,2s}^{\rm R}+\delta V_{2r,2s-1}^{\rm I}\delta U_{2r-1,2s}^{\rm I}\right],\nonumber
\end{align}
where we abbreviated $a=-4\sin\alpha_r\sin\alpha_s$ and $b=-4\cos\alpha_r\cos\alpha_s$. 

The contribution to the metric tensor is a block matrix with elements
\begin{equation}
\begin{pmatrix}
 4 \tau_0 & 0  & 0 & 0 & a \tau_z & 0 & 0 & b \tau_0  \\ 
                                           0 & 4 \tau_0 & 0  & 0 & 0 & a \tau_z & -b \tau_0 & 0 \\ 
                                           0 & 0 & 4 \tau_0 & 0  & 0 & -b \tau_0 & a \tau_z & 0 \\ 
                                           0 & 0 & 0 & 4 \tau_0 & b \tau_0  & 0 & 0 & a \tau_z  \\ 
                                           a\tau_z & 0 & 0 & b\tau_0 &  4 \tau_0 & 0 & 0 & 0   \\ 
                                           0 & a\tau_z & -b \tau_0 & 0 & 0 &  4 \tau_0 & 0 & 0  \\ 
                                           0 & -b \tau_0 & a\tau_z & 0 & 0 & 0 &  4 \tau_0 & 0  \\ 
                                           b \tau_0 & 0 & 0 & a\tau_z & 0 & 0 & 0 &  4 \tau_0  
\end{pmatrix},\nonumber
\end{equation}
where the Pauli matrix $\tau_z$ and the $2\times 2$ unit matrix $\tau_0$ were introduced to account for real and imaginary parts in a compact way. The determinant of this matrix is $(\sin^2 \alpha_r-\sin^2 \alpha_s)^{8}$, hence the contribution to the Jacobian from the off-diagonal matrix elements is 
\begin{equation}
J_{\text{off-diagonal}}=\prod_{r<s=1}^M \left| \sin^2 \alpha_r-\sin^2 \alpha_s\right|^4.
\end{equation} 

Next we consider the diagonal blocks. We choose as independent parameters 
\begin{align}
&w_1=-i(\delta V_{2m,2m}-\delta U_{2m-1,2m-1}),\nonumber\\
&w_2=-i(\delta V_{2m-1,2m-1}-\delta U_{2m,2m}),\nonumber\\
&w_3=-i(\delta V_{2m,2m}+\delta U_{2m,2m}),\nonumber\\
&w_4=\delta V_{2m-1,2m}+\delta U_{2m,2m-1}.\nonumber
\end{align}
These are in total $5M$ real parameters. (Since $w_1,w_2,w_3$ are real numbers they contribute only $M$ parameters each.) The contribution to $\text{Tr}\,\delta r \delta r^\dag$ is 
\begin{equation}
w_1^2+w_2^2+2 (w_2 w_3-w_1w_3-w_1w_2+w_{3}^{2})\sin^2\alpha_m +2 w_4^2,\nonumber
\end{equation}
and the contribution to the metric tensor is the block matrix
\begin{equation}
\begin{pmatrix}
2&0&0&0\\
0& 1 & -\sin^2 \alpha_m & -\sin^2 \alpha_m   \\ 
0&                                         -\sin^2 \alpha_m & 1 & \sin^2 \alpha_m \\ 
0&                                          -\sin^2 \alpha_m & \sin^2 \alpha_m & 2\sin^2 \alpha_m  
\end{pmatrix},\nonumber
\end{equation}
with determinant $2( \sin \alpha_{m}\cos \alpha_m)^{2}$. Hence the contribution to the Jacobian from the diagonal matrix elements is
\begin{equation}
J_{\rm diagonal}=\prod_{m=1}^{M}|\sin \alpha_{m}\cos \alpha_m|.
\end{equation}

The number of independent parameters that we have accounted for totals to $8M^2-2M$, which should equal the number of degrees of freedom of a matrix in class D. The matrix space in class D is isomorphic to the group of $2N\times 2N$ orthogonal matrices,\cite{Alt97} which indeed has $N(2N-1)=8M^2-2M$ degrees of freedom.

Gathering all terms that contribute to the Jacobian in Eq.\ \eqref{Adistribution}, we obtain the probability distribution
\begin{equation}
P(\{\alpha_n\})\propto \prod_{r<s=1}^M \left| \sin^2 \alpha_r-\sin^2 \alpha_s\right|^4  \prod_{m=1}^M \left| \sin \alpha_{m}\cos \alpha_m\right|.
\end{equation}
The integration $\int dp_{i}$ over the degrees of freedom of the eigenvector matrices only contributes a prefactor, which can be absorbed in the proportionality constant. Upon transformation to the Andreev reflection eigenvalues $R_n=\sin^2\alpha_n$, we arrive at the result \eqref{PCREa} stated in the main text.

\subsection{Class DIII (ensemble T-CRE)}
\label{appA_DIII}

For the treatment of class DIII it is useful to notice the similarity of the polar decomposition of $i\Sigma_y r$ given in Eq.\ (\ref{iSigmayr}) to the one of $r$ in class D given in Eq. (\ref{polarD}). Since $\delta (i\Sigma_y r)=\delta r$ all the equations up to Eq. (\ref{Tr3}) derived for class D also hold for class DIII, upon replacement $U\mapsto\Omega$ and $V\mapsto\Omega^*$. (As before, we only give the detailed derivation for $Q=1$.) The expressions for the traces $T_{1}$ and $T_{3}$ then simplify to
\begin{widetext}
\begin{align}
T_1={}&\sum_{r<s=1}^M \left\{ (\cos \alpha_{r}+\cos \alpha_s)^2 \left[  |\delta \Omega_{2r,2s}-\delta \Omega^*_{2r-1,2s-1}|^2+ |\delta \Omega_{2r-1,2s}+\delta \Omega^*_{2r,2s-1}|^2\right]\right. \nonumber\\
&+\left.(\cos \alpha_{r}-\cos \alpha_s)^2 \left[  |\delta \Omega_{2r,2s}+\delta \Omega^*_{2r-1,2s-1}|^2+ |\delta \Omega_{2r-1,2s}-\delta \Omega^*_{2r,2s-1}|^2\right]\right\}\nonumber\\
&+2\sum_m^M \cos^2\alpha_m |\delta \Omega_{2m,2m}+\delta \Omega_{2m-1,2m-1}|^2,\\
T_3={}&2\sum_{r<s=1}^M \sum_{k,l=0}^1 (\sin \alpha_{r}-\sin \alpha_s)^2  |\delta \Omega_{2r-k,2s-l}|^2.
\end{align}
\end{widetext}

For the off-diagonal blocks we choose
\begin{align}
\delta \Omega_{2r,2s},\;\;\delta \Omega_{2r-1,2s},\;\;\delta \Omega_{2r,2s-1},\;\;\delta \Omega_{2r-1,2s-1},\nonumber
\end{align}
with  $1\le r<s\le M$, as the independent real parameters (a total of $4M^2-4M$). The contribution to $\text{Tr}\,\delta r \delta r^\dag$ for given values of $r$ and $s$ is
\begin{align}
&c \sum_{k=2r-1}^{2r}\sum_{l=2s-1}^{2s}\left[(\delta \Omega^{\rm R}_{kl})^2+(\delta \Omega^{\rm I}_{kl})^2\right]\nonumber\\
&+2d \left[ \delta \Omega_{2r-1,2s}^{\rm R}\delta \Omega_{2r,2s-1}^{\rm R}-\delta \Omega_{2r-1,2s}^{\rm I}\delta \Omega_{2r,2s-1}^{\rm I}\right.\nonumber\\
&\quad\left.-\delta \Omega_{2r-1,2s-1}^{\rm R}\delta \Omega_{2r,2s}^{\rm R}+\delta \Omega_{2r-1,2s-1}^{\rm I}\delta \Omega_{2r,2s}^{\rm I}\right],\nonumber
\end{align}
with $c=4-4\sin \alpha_r\sin \alpha_s$ and $d=4\cos\alpha_r\cos\alpha_s$. The contribution to the metric tensor is a block matrix with elements
\begin{equation}
\begin{pmatrix}
c \tau_0 & 0  & 0 & -d \tau_z   \\ 
                                           0 & c \tau_0 & d\tau_z  & 0   \\ 
                                           0 & d\tau_z & c \tau_0 & 0   \\ 
                                           -d \tau_z & 0 & 0 & c\tau_0    
\end{pmatrix},\nonumber 
\end{equation}
with determinant $(\sin \alpha_r-\sin \alpha_s)^8$. The contribution to the Jacobian is
\begin{equation}
J_{\text{off-diagonal}}=\prod_{r<s=1}^M \left| \sin \alpha_r-\sin \alpha_s\right|^4.
\end{equation} 

The diagonal blocks have $M$ independent degrees of freedom,
\begin{equation}
\delta \Omega_{2m\,2m}+\delta \Omega_{2m-1\,2m-1},\nonumber
\end{equation}
which contribute to the Jacobian a factor
\begin{equation}
J_{\text{diagonal}}=\prod_{m=1}^M \left| \cos \alpha_m\right|.
\end{equation}

The total number of independent parameters (including also the $M$ degrees of freedom from the $\alpha_{n}$'s) is then $4M^2-2M$. This agrees with the number of degrees of freedom of the matrix space $O(2N)/U(N)$ in class DIII.\cite{Alt97} 

The distribution of the $\alpha_{n}$'s results from the product of $J_{\text{off-diagonal}}$ and $J_{\text{diagonal}}$,
\begin{equation}
P(\{\alpha_n\})\propto \prod_{r<s=1}^M \left| \sin \alpha_r-\sin \alpha_s\right|^4  \prod_{m=1}^M \left| \cos \alpha_m\right|.
\end{equation}
Transformation to $\xi_n=\sin\alpha_n$ gives the expression \eqref{PTCREa} in the main text.

\section{Proof of the topological-charge theorem for the circular ensembles}
\label{AppB}

The theorem we wish to prove states that the $p$-th cumulant of the conductance in the $N$-mode circular ensemble is independent of the topological charge $Q$ for $p<N/d$, with $d=1$ in the CRE and $d=2$ in the T-CRE. 

We start from the definition \eqref{BTK} of the conductance, which we rewrite as
\begin{equation}
G/G_{0}=\tfrac{1}{2}{\rm Tr}\,[1-r^{\dagger}\tau_{z}r(1+\tau_{z})],\;\;\tau_{z}=\begin{pmatrix}
1&0\\
0&-1
\end{pmatrix}.\label{Gtauz}
\end{equation}
The reflection matrix $r$ is a $2N\times 2N$ unitary matrix, satisfying the particle-hole symmetry relation \eqref{ehsymmetry}, which we rewrite as
\begin{equation}
r=\tau_{x}r^{\ast}\tau_{x},\;\;\tau_{x}=\begin{pmatrix}
0&1\\
1&0
\end{pmatrix}.\label{ehtaux}
\end{equation}
This equation implies that ${\rm Tr}\,r^{\dagger}\tau_{z}r=0$, hence Eq.\ \eqref{Gtauz} reduces to
\begin{equation}
G/G_{0}=\tfrac{1}{2}{\rm Tr}\,[1-r^{\dagger}\tau_{z}r\tau_{z})].\label{Gtauz2}
\end{equation}

The $p$-th cumulant of $G$ contains only averages $m_{q}=\langle ({\rm Tr}\,r^{\dagger}\tau_{z}r\tau_{z})^{q}\rangle$ with $q\leq p$, hence to prove the theorem is it sufficient to prove that $m_{p}$ is independent of $Q$ for $p<N/d$. 

We first do this for the CRE. Then the average $m_{p}$ can be written as
\begin{equation}
m_{p}=\int d\mu(r)\,\bigl({\rm Tr}\,r^{\dagger}\tau_{z}r\tau_{z}\bigr)^{p}\tfrac{1}{2}(1+Q\,{\rm Det}\,r),\label{mpCRE}
\end{equation}
where $d\mu(r)$ is the invariant measure of class D. The defining property of this measure is that $d\mu(Ur)=d\mu(rU)=d\mu(r)$ for any $2N\times 2N$ unitary matrix $U$ that satisfies $U=\tau_{x}U^{\ast}\tau_{x}$. What we seek to prove, therefore, is that
\begin{equation}
\int d\mu(r)\,\bigl({\rm Tr}\,r^{\dagger}\tau_{z}r\tau_{z}\bigr)^{p}\,{\rm Det}\,r=0\;\;{\rm if\;\;}p<N.\label{Gtauz3}
\end{equation}

We decompose
\begin{equation}
\tau_{z}=\sum_{n=1}^{N}\tau^{(n)},\;\;\tau^{(n)}_{kl}=\delta_{k,l}(\delta_{k,n}-\delta_{k,n+N})\label{taundef}
\end{equation}
and apply this decomposition to one of the $\tau_{z}$'s in Eq.\ \eqref{Gtauz3},
\begin{align}
&\bigl({\rm Tr}\,r^{\dagger}\tau_{z}r\tau_{z}\bigr)^{p}=\sum_{p_{1}=0}^{N}\sum_{p_{2}=0}^{N}\cdots\sum_{p_{N}=0}^{N}\frac{p!}{p_{1}!p_{2}!\cdots p_{N}!}\nonumber\\
&\quad\times\delta_{p,p_{1}+p_{2}+\cdots +p_{N}}\prod_{n=1}^{N}\left({\rm Tr}\,r^{\dagger}\tau^{(n)}r\tau_{z}\right)^{p_{n}}.\label{Gtauz4}
\end{align}

Consider one of the terms
\begin{equation}
{\cal M}=\int d\mu(r)\,\prod_{n=1}^{N}\left({\rm Tr}\,r^{\dagger}\tau^{(n)}r\tau_{z}\right)^{p_{n}}\,{\rm Det}\, r.\label{calMdef}
\end{equation}
If $p<N$, there is at least one index $n_{0}\in\{1,2,\ldots N\}$ such that $p_{n_{0}}=0$. Transform $r\mapsto U^{(n_{0})}r$, with
\begin{equation}
U^{(n_{0})}_{kl}=\left\{\begin{array}{ll}
\delta_{k,l}&{\rm if}\;\;k\neq n_{0},n_{0}+N,\\
\delta_{l,n_{0}+N}&{\rm if}\;\;k=n_{0},\\
\delta_{l,n_{0}}&{\rm if}\;\;k=n_{0}+N,
\end{array}\right.\label{Un0def}
\end{equation}
a real, symmetric, unitary matrix which commutes with $\tau_{x}$. This transformation does not change the invariant measure, $d\mu(U^{(n_{0})}r)=d\mu(r)$, while the integrand transforms to
\begin{align}
{\cal M}={}&\int d\mu(r)\,\prod_{n=1}^{N}\left({\rm Tr}\,r^{\dagger}U^{(n_{0})}\tau^{(n)}U^{(n_{0})}r\tau_{z}\right)^{p_{n}}\nonumber\\
&\times{\rm Det}\, U^{(n_{0})}r\nonumber\\
=&-\int d\mu(r)\,\prod_{n=1}^{N}\left({\rm Tr}\,r^{\dagger}\tau^{(n)}r\tau_{z}\right)^{p_{n}}\,{\rm Det}\, r\nonumber\\
=&-{\cal M},\label{Gtauz5}
\end{align}
since ${\rm Det}\,U^{(n_{0})}=-1$ and $U^{(n_{0})}$ commutes with $\tau^{(n)}$ for $n\neq n_{0}$, while $p_{n_{0}}=0$. Hence ${\cal M}=0$.

This completes the proof for the CRE. For the T-CRE, we seek to prove that
\begin{equation}
\int d\mu(r)\,\bigl({\rm Tr}\,r^{\dagger}\tau_{z}r\tau_{z}\bigr)^{p}\,{\rm Pf}\,i\Sigma_{y}r=0\;\;{\rm if\;\;}p<N/2,\label{toproveTCRE}
\end{equation}
where now $d\mu(r)$ is the invariant measure of class DIII. The invariance property reads $d\mu(\Sigma_{y}U^{\rm T}\Sigma_{y}rU)=d\mu(r)$ for any $2N\times 2N$ unitary matrix $U$ that satisfies $U=\tau_{x}U^{\ast}\tau_{x}$. Since $\tau_{z}$ and $\Sigma_{y}$ commute, we may rewrite Eq.\ \eqref{toproveTCRE} as
\begin{equation}
\int d\mu(r)\,\bigl({\rm Tr}\,r^{\dagger}\Sigma_{y}\tau_{z}\Sigma_{y}r\tau_{z}\bigr)^{p}\,{\rm Pf}\,i\Sigma_{y}r=0\;\;{\rm if\;\;}p<N/2.\label{toproveTCRE2}
\end{equation}

Substitute the decomposition \eqref{taundef} in both the $\tau_{z}$'s,
\begin{align}
&\bigl({\rm Tr}\,r^{\dagger}\Sigma_{y}\tau_{z}\Sigma_{y}r\tau_{z}\bigr)^{p}=\sum_{p_{11}=0}^{N}\cdots\sum_{p_{NN}=0}^{N}\frac{p!}{\prod_{n,m}p_{nm}!}\nonumber\\
&\quad\times\delta_{p,\sum_{n,m}p_{nm}}\prod_{n,m=1}^{N}\left({\rm Tr}\,r^{\dagger}\Sigma_{y}\tau^{(n)}\Sigma_{y}r\tau^{(m)}\right)^{p_{nm}}.\label{expandboth}
\end{align}
Consider one of the terms
\begin{align}
{\cal M}={}&\int d\mu(r)\,\prod_{n,m=1}^{N}\left({\rm Tr}\,r^{\dagger}\Sigma_{y}\tau^{(n)}\Sigma_{y}r\tau^{(m)}\right)^{p_{nm}}\nonumber\\
&\times{\rm Pf}\, i\Sigma_{y}r.\label{oneterm}
\end{align}
If $p<N/2$, there is at least one index $n_{0}\in\{1,2,\ldots N\}$ such that $p_{n_{0}m}=0$ and $p_{nn_{0}}=0$ for each $n,m\in\{1,2,\ldots N\}$. Transform $r\mapsto \Sigma_{y}U^{(n_{0})}\Sigma_{y}rU^{(n_{0})}$, with $U^{(n_{0})}$ defined in Eq.\ \eqref{Un0def}. This transformation does not change the invariant measure, so the integral transforms to
\begin{align}
{\cal M}={}&\int d\mu(r)\,\prod_{n,m=1}^{N}\left({\rm Tr}\,r^{\dagger}\Sigma_{y}U^{(n_{0})}\tau^{(n)}U^{(n_{0})}\Sigma_{y}r\right.\nonumber\\
&\times\left.U^{(n_{0})}\tau^{(m)}U^{(n_{0})}\right)^{p_{nm}}\,{\rm Pf}\,\bigl(U^{(n_{0})}i\Sigma_{y}rU^{(n_{0})}\bigr)\nonumber\\
=&-\int d\mu(r)\,\prod_{n,m=1}^{N}\left({\rm Tr}\,r^{\dagger}\Sigma_{y}\tau^{(n)}\Sigma_{y}r\tau^{(m)}\right)^{p_{nm}}\nonumber\\
&\times\,{\rm Pf}\, i\Sigma_{y}r
=-{\cal M},\label{calMis0}
\end{align}
where we have used that ${\rm Pf}\,XYX^{\rm T}=({\rm Det}\,X)({\rm Pf}\,Y)$. Hence ${\cal M}=0$ and we have completed the proof for the T-CRE.


\begin{thebibliography}{99}
\bibitem{Blu90} R. Bl\"{u}mel and U. Smilansky, Phys. Rev. Lett. \textbf{64}, 241 (1990).
\bibitem{Dys62} F. J. Dyson, J. Math. Phys. \textbf{3}, 140 (1962).
\bibitem{RMTbook} \textit{Handbook on Random Matrix Theory}, edited by G. Akemann, J. Baik, and P. Di Francesco. The circular ensembles are reviewed in Chapters by: M. R. Zirnbauer, arXiv:1001.0722, Y. V. Fyodorov and D. V. Savin, arXiv:1003.0702; C. W. J. Beenakker, arXiv:0904.1432.
\bibitem{Alt97} A. Altland and M. R. Zirnbauer, Phys. Rev. B \textbf{55}, 1142 (1997).
\bibitem{Dah10} J. P. Dahlhaus, B. B\'{e}ri, and C. W. J. Beenakker, Phys. Rev. B \textbf{82}, 014536 (2010).
\bibitem{Kit01} A. Yu. Kitaev Phys. Usp. \textbf{44} (suppl.), 131 (2001); arXiv:0901.2686.
\bibitem{Qi08} X.-L. Qi, T. L. Hughes, and S.-C. Zhang, Phys. Rev. B \textbf{78}, 195424 (2008);  X.-L. Qi and S.-C. Zhang, arXiv:1008.2026.
\bibitem{Sch08} A. P. Schnyder, S. Ryu, A. Furusaki, and A. W. W. Ludwig, Phys. Rev. B \textbf{78} 195125 (2008); S. Ryu, A. P. Schnyder, A. Furusaki, and A. W. W. Ludwig, New J. Phys. \textbf{12}, 065010 (2010).
\bibitem{Fu09} L. Fu and C. L. Kane, Phys. Rev. B 79, 161408(R) (2009); M. Z. Hasan and C. L. Kane, Rev. Mod. Phys. \textbf{82}, 3045 (2010).
\bibitem{Lut10a} R. M. Lutchyn, J. D. Sau, and S. Das Sarma, Phys. Rev. Lett. \textbf{105}, 077001 (2010).
\bibitem{Ore10} Y. Oreg, G. Refael, and F. von Oppen, Phys. Rev. Lett. \textbf{105}, 177002 (2010).
\bibitem{Wim10} M. Wimmer, A. R. Akhmerov, M. V. Medvedyeva, J. Tworzyd{\l}o, and C. W. J. Beenakker, Phys. Rev. Lett. \textbf{105}, 046803 (2010).
\bibitem{Pot10} A. C. Potter and P. A. Lee, Phys. Rev. Lett. \textbf{105}, 227003 (2010).
\bibitem{Lut10b} R. M. Lutchyn, T. Stanescu, and S. Das Sarma, arXiv:1008.0629.
\bibitem{Duc10} M. Duckheim and P. W. Brouwer, arXiv:1011.5839.
\bibitem{Law09} K. T. Law, P. A. Lee, and T. K. Ng, Phys. Rev. Lett. \textbf{103}, 237001 (2009).
\bibitem{Fle10} K. Flensberg, Phys. Rev. B \textbf{82}, 180516(R) (2010).
\bibitem{Ber09b} B. B\'{e}ri, J. N. Kupferschmidt, C. W. J. Beenakker, and P. W. Brouwer, Phys. Rev. B \textbf{79}, 024517 (2009).
\bibitem{Wim11} M. Wimmer, A. R. Akhmerov, J. P.  Dahlhaus, and C. W. J. Beenakker, arXiv:1101.5795.
\bibitem{Kha10} I. M. Khaymovich, N. M. Chtchelkatchev, I. A. Shereshevskii, and A. S. Mel'nikov, EPL \textbf{91}, 17005 (2010).
\bibitem{Blo82} G. E. Blonder, M. Tinkham, and T. M. Klapwijk, Phys. Rev. B \textbf{25}, 4515 (1982).
\bibitem{Bar08} J. H. Bardarson, J. Phys. A \textbf{41}, 405203 (2008).
\bibitem{Ber09a} B. B\'{e}ri, Phys. Rev. B \textbf{79}, 245315 (2009).
\bibitem{Boc00} M. Bocquet, D. Serban, and M. Zirnbauer, Nucl. Phys. B \textbf{578}, 628 (2000).
\bibitem{Mer02} F. Merz and J. T. Chalker, Phys. Rev. B \textbf{65}, 054425 (2002).
\bibitem{Akh10} A. R. Akhmerov, J. P. Dahlhaus, F. Hassler, M. Wimmer, and C. W. J. Beenakker, Phys. Rev. Lett. \textbf{106}, 057001 (2011).
\bibitem{Ful11} I. C. Fulga, F. Hassler, A. R. Akhmerov, and C. W. J. Beenakker, arXiv:1101.1749.
\bibitem{Bee97} C. W. J. Beenakker, Rev. Mod. Phys. \textbf{69}, 731 (1997).
\bibitem{For10} P. J. Forrester, \textit{Log-Gases and Random Matrices} (Princeton University Press, Princeton, 2010).
\bibitem{Bee93} C. W. J. Beenakker, Phys. Rev. B \textbf{47}, 15763 (1993).
\bibitem{Wim09} M. Wimmer and K. Richter, J. Comput. Phys. \textbf{228}, 8548 (2009).
\end{thebibliography}
\end{document}